# A PBPO$^+$ Graph Rewriting Tutorial

Roy Overbeek      Jörg Endrullis

Vrije Universiteit Amsterdam
Amsterdam, The Netherlands

`r.overbeek@vu.nl`      `j.endrullis@vu.nl`

We provide a tutorial introduction to the algebraic graph rewriting formalism PBPO$^+$. We show how PBPO$^+$ can be obtained by composing a few simple building blocks, and model the reduction rules for binary decision diagrams as an example. Along the way, we comment on how alternative design decisions lead to related formalisms in the literature, such as DPO. We close with a detailed comparison with Bauderon's double pullback approach.

## 1 Introduction

Graphs are useful data structures for virtually every field of computer science. And they come in many varieties: they can be directed or undirected, may carry labels or attributes, the edges can be hyperedges, and so on. Computation on these varieties is however similar, and can roughly be described as the stepwise transformation of subgraphs, induced by a set of transformation rules. For this reason, it is helpful if the transformation framework can be defined in a general way, without having to commit to a particular graph notion. Ehrig et al. [12] first managed to do so in the 1970s, by specifying a framework (namely, Double Pushout (DPO) rewriting) in the language of category theory. Their seminal work laid the foundation for the field of algebraic graph rewriting [10], in which a variety of categorical graph transformation frameworks have been proposed and studied. Apart from DPO, well-known frameworks in this field include SPO [17], double pullback rewriting [3, 4], SqPO [8], AGREE [6] and PBPO [7].

Why this variety in frameworks? At least three factors are relevant:

- *Scope*: not all frameworks are well-defined for exactly the same classes of graphs.
- *Replacement in unknown context*: suppose that a rewrite rule $\rho$ specifies the deletion of an arbitrary vertex in a graph. We select a vertex $v$ of a graph, and see that it has edges incident to it. Is $\rho$ at all applicable at $v$? And if $\rho$ is applicable, then what should happen to the incident edges? Should they be deleted? Redirected? Different policies are possible. Some frameworks fix the policy, whereas others effectively allow the user to define the policy on a rule-by-rule basis.
- *Duplication*: some frameworks provide support for duplicating elements, whereas others do not.

One could regret the fact that the pursuit for abstraction and generalization has nonetheless led to such a fragmentation of formalisms. In an attempt to provide some level of unification, we have recently proposed PBPO$^+$ [18] (short for *Pullback-Pushout with strong matching*; a modification of PBPO by Corradini et al. [7]) and shown that in the abstract setting of quasitoposes, PBPO$^+$ can define a strict superset of the rewrite relations definable by DPO, SqPO, AGREE and PBPO. Quasitoposes include various important graph-like categories such as the categories of labeled multigraphs, hypergraphs, safely marked Petri nets and simple graphs; as well as various categories that are not graph-like. Moreover, our proofs are constructive in that they show how rewrite systems of the mentioned formalisms can be naturally encoded into PBPO$^+$. As a consequence, theory, methods and tools developed for PBPO$^+$ extend to these other formalisms for this setting. We posit that this fact makes PBPO$^+$ worth studying.

In this tutorial, we introduce PBPO$^+$ in a stepwise manner, starting from minimal preliminaries. In particular, we do not assume an understanding of category theory or related formalisms. Instead, we introduce two toy formalisms, ToyPO (Section 2) and ToyPB (Section 3), and show how PBPO$^+$ can be understood as a combination of the two, using the category of unlabeled graphs as our setting (Section 4). Then, we switch to the setting of labeled graphs, and show how binary decision diagram reduction can be modeled (Section 5). After, we zoom out and clarify the relationship between PBPO$^+$ and the double pullback rewriting approach by Bauderon [3], which bears some similarities to PBPO$^+$ (Section 6). We close with a short overview of tutorials for related formalisms (Section 7).

*Preliminaries.* We write $f \circ g$ to denote function composition $((f \circ g)(x) = f(g(x)))$. By a *graph* $G$ we mean an *unlabeled directed multigraph*, i.e., $G = (V, E, s, t)$, where $V$ is a set of *vertices* (or *nodes*), $E$ a set of *edges*, $s : E \to V$ a *source* function and $t : E \to V$ a *target* function. A *graph homomorphism* $\phi : G \to H$ from graph $G$ to graph $H$ is a pair of functions $\phi_V : V_G \to V_H$ and $\phi_E : E_G \to E_H$ satisfying $\phi_V \circ s_G = s_H \circ \phi_E$ and $\phi_V \circ t_G = t_H \circ \phi_E$. We use $\rightarrowtail$ to denote injective homomorphisms, and write $A \cong B$ to denote that $A$ and $B$ are isomorphic.

*A note on vocabulary.* A category consists of *objects* and *morphisms* between them. In our graph setting, these instantiate to graphs, and graph homomorphisms between graphs, respectively. We state some concepts (such as pushouts and pullbacks) categorically in this paper, but readers unacquainted with category theory may safely read "graph" for "object" and "graph homomorphism" for "morphism".

## 2 ToyPO

We start by defining a toy graph rewrite formalism called ToyPO (short for ToyPushout). This formalism computes rewrite steps using a single pushout construction. It allows identifying and adding elements.

**Definition 1** (ToyPO Rule). A *ToyPO rule* is a morphism $\rho : L \to R$. $L$ and $R$ are called *patterns*.

*Example* 1. The rule $\rho : L \to R$ depicted by 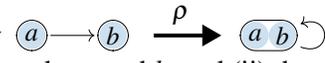 can be described as specifying (i) the identification of two connected nodes $a$ and $b$, and (ii) the addition of a node $c$.

*Notation* 1 (Visual Notation). Our notation for graphs and graph homomorphisms can be described formally. A vertex is a non-empty set $\{x_1, \ldots, x_n\}$ represented by a box $(x_1 \cdots x_n)$, and each morphism $\phi = (\phi_V, \phi_E) : G \to G'$ between depicted graphs $G$ and $G'$ is the unique morphism satisfying $S \subseteq \phi(S)$ for all $S \in V_G$. For instance, for nodes $\{a\}$ and $\{b\}$ in the left hand side of $\rho$ (Example 1), $\rho(\{a\}) = \rho(\{b\}) = \{a,b\}$ of the right hand side. For notational convenience, will use examples that ensure uniqueness of each $\phi$ (in particular, we ensure that the mapping of edges $\phi_E$ is uniquely determined by the mapping of nodes $\phi_V$). Colors are purely supplementary.

**Definition 2** (Match). A *match* for a ToyPO rule $\rho : L \to R$ in a *host graph* $G$ is an injective morphism $m : L \rightarrowtail G$. The image $m(L) \subseteq G$ is an *occurrence* of $L$ in $G$.

*Example* 2. The injective morphism $m : L \rightarrowtail G$ depicted by is a match for rule $\rho$ of Example 1 in the host graph $G$. Three other matches are possible, with images $d \to a$, $e \to f$ and $f \to e$, respectively. The homomorphism that has $f \to f$ as its image is not a match, because it is not injective.

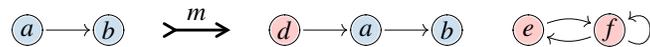

*Remark* 1. For this tutorial, we require injectivity of $m$ for three reasons. First, it makes matches easier to understand. Second, it yields a strictly more expressive formalism (cf. Habel et al. [13]). Third, in practice, matches often are required to be injective.

Given rule $\rho : L \to R$ and match $m : L \rightarrowtail G$ of Examples 1 and 2, respectively, arguably the most reasonable result of the rewrite step is the graph because this graph can be understood as the most natural solution in which $m(L)$ is replaced by $R$ in $G$. More specifically, no elements of $G - m(L)$ are deleted, duplicated or identified; and no elements other than those in $R - \rho(L)$ are added. Remarkably, this informal notion of a natural solution can be defined formally and abstractly using the language of category theory; that is, without making any reference whatsoever to graphs or graph homomorphisms.

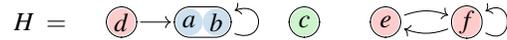

**Definition 3** (Pushout [20, 2]). The *pushout* of a *span* $G \xleftarrow{m} L \xrightarrow{\rho} R$ is a *cospan* $\sigma = G \xrightarrow{i_G} H \xleftarrow{i_R} R$ such that

1. $\sigma$ is a candidate solution: $i_G \circ m = i_R \circ \rho$; and

2. $\sigma$ is the universal solution: for any cospan $G \xrightarrow{i_G'} H' \xleftarrow{i_R'} R$ that satisfies $i_G' \circ m = i_R' \circ \rho$, there exists a *unique* morphism $x : H \to H'$ such that $i_G' = x \circ i_G$ and $i_R' = x \circ i_R$.

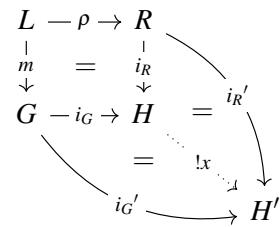

Figure 1: Pushout.

Both requirements are conveniently visualized in the commuting diagram depicted in Figure 1.

Pushouts are unique up to isomorphism if they exist (and they always do in the category of graphs). A useful intuitive description of a pushout is that of a *gluing construction*, because $H$ can be thought of as the result of gluing $G$ and $R$ along shared interface $L$. This view is especially intuitive if $m$ is injective, which is always the case in this tutorial.

*Exercise* 1. For span $G \xleftarrow{m} L \xrightarrow{\rho} R$ given by Examples 1 and 2, verify that

1. the following pushout is indeed a candidate solution (i.e., satisfies condition 1 of Definition 3):

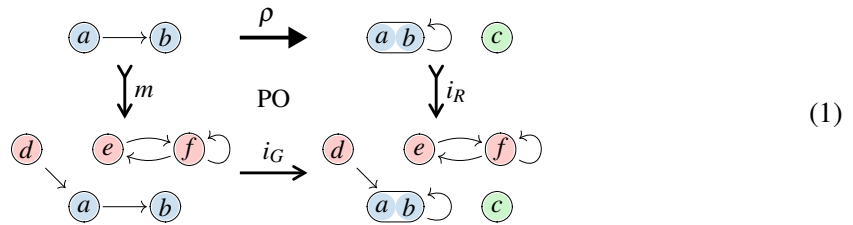

(1)

2. the following graphs (with the obvious choices for the cospan morphisms $i_G, i_R$) constitute two other candidate solutions:

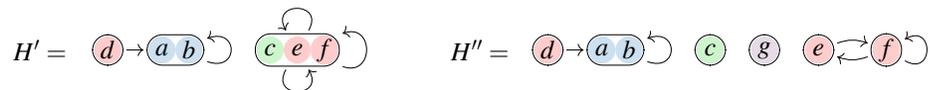

3. $H$ satisfies condition 2 of Definition 3 specifically for competing candidates $H'$ and $H''$; and
4. $H'$ and $H''$ both fail condition 2 of Definition 3 because
    (a) no suitable witness $x : H' \to H$ exists (the identification of elements cannot be undone); and
    (b) a suitable witness $x : H'' \to H$ does not exist uniquely (node $g$ can be mapped freely).

For a set theoretic description of the pushout, one may consult König et al. [16]. For this paper, however, it suffices if the reader understands the pushout intuitively as a gluing construction, and that it can be used for the specification of identification and addition of elements.

We may now define the notion of a ToyPO rewrite step.

**Definition 4** (ToyPO Rewrite Step). A ToyPO rule $\rho : L \to R$ and match $m : L \rightarrowtail G$ induce a *ToyPO rewrite step* $G \Rightarrow_{\text{ToyPO}}^{\rho,m} H$ if there exists a pushout of the form

$$\begin{array}{ccc} L & \xrightarrow{\rho} & R \\ m \downarrow & \text{PO} & \downarrow i_R \\ G & \xrightarrow{i_G} & H \end{array}.$$

## 3 ToyPB

Next, we would like to specify the duplication and deletion of elements. This is slightly more challenging, and allows for a larger variety of solutions.

For some historical context, and to illustrate the solution space, consider the approach where we invert the procedure described in Section 2. For example, we read rule $\rho : L \to R$ of Example 1 right-to-left, suggestively writing the type signature as $\rho : R \leftarrow L$. Read in this way, $\rho$ intuitively specifies the duplication of node *ab* and the deletion of node *c*. When we find a match $m : R \rightarrowtail G$, we try to show that $G$ can be understood as the result of a suitable gluing. That is, we try to find morphisms $c_1 : L \to H$ and $c_2 : H \to G$ such that the diagram

$$\begin{array}{ccc} R & \xleftarrow{\rho} & L \\ m \downarrow & \text{PO} & \downarrow c_1 \\ G & \xleftarrow{c_2} & H \end{array}$$

is a pushout square. Morphisms $c_1$ and $c_2$ together constitute a *pushout complement* for $\rho$ and $m$.

*Example* 3. The pushout square of Diagram (1) in Exercise 1 can be read as an inverted application of $\rho$, by reading morphism $i_R$ as the match, the bottom right graph as the host graph, and the bottom left graph as the result graph.

The inverse approach has two particular caveats:

1. *Existence of pushout complements*: Consider a rule $\rho$ that deletes a node $a$ (the domain of $\rho$ is the empty graph), together with a match $m$, as depicted in:

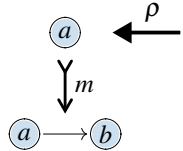

Even though all pushouts exist in the category of graphs, and even though we have found a match $m$, a pushout complement does not exist for $(m, \rho)$: the depicted edge can never be obtained through a gluing around an empty interface. The more general implication of this observation is that we are only able to delete nodes if they do not leave any edges "dangling" (called the *gluing condition*). This can be considered a pleasant safety feature, but also a limitation: perhaps we would prefer to simply delete any incident edges (as a general principle), or even be able to specify more fine-grained control on the level of the rule itself.

2. *Uniqueness of pushout complements*: More importantly, unlike pushouts, pushout complements need not be unique: the square

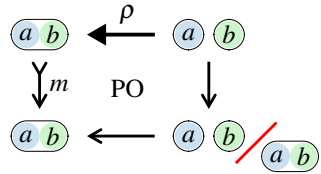

shows that two pushout complements exist, one of which does not model duplication (similarly, there exists another pushout complement for the pushout square of Diagram (1)). This is problematic if it is desired that rewrite step results are uniquely determined by the rule and match. In the category of graphs, uniqueness of pushout complements can at least be ensured by restricting rules $\rho$ to injective morphisms (thus allowing deletion, but not duplication of elements). However, in other categories, such as the category of simple graphs, this is not generally sufficient.

These two caveats notwithstanding, the most dominant graph rewriting method, called the Double Pushout (DPO) approach, combines the pushout complement approach (with the injectivity requirement on $\rho$) with the ToyPO approach. In doing so, it enables the specification of rewrite steps with deletion, identification and addition features, for matches $m$ that satisfy the gluing condition.

**Definition 5** (DPO Rewriting [12]). A *DPO rewrite rule* $\rho$ is a span $L \hookleftarrow K \xrightarrow{r} R$. A diagram

$$
\begin{array}{ccccc}
L & \leftarrow l \!\!\rightarrowtail & K & \longrightarrow r \longrightarrow & R \\
\rotatebox{90}{$\hookrightarrow$}\!m\!\downarrow & \text{PO} & \downarrow & \text{PO} & \downarrow \\
G_L & \longleftarrow & G_K & \longrightarrow & G_R
\end{array}
$$

defines a *DPO rewrite step* $G_L \Rightarrow_{\text{DPO}}^{\rho,m} G_R$, i.e., a step from $G_L$ to $G_R$ using rule $\rho$ and match $m : L \rightarrowtail G_L$.

Alternatives to the DPO approach avoid the construction of pushout complements. For instance, the Single Pushout (SPO) approach [17] relies on a single pushout construction, but uses partial graph homomorphisms instead of total morphisms, in order to specify deletion. In this approach, the gluing condition no longer needs to be checked either: all edges incident to a removed vertex are simply deleted. As another example, the Sesqui-Pushout (SqPO) approach [8] replaces the first PO square of DPO by what is called a *final pullback complement* square. This square allows duplication with deterministic behavior, and like SPO, deletes any edges that would be left dangling.

For our proposal, called ToyPB (short for ToyPullback), we consider a perspective that is *dual* to the ToyPO approach, rather than its inverse. First, instead of trying to find an occurrence of some pattern $L$ in a graph $G$ (through a match $m : L \rightarrowtail G$), we try to find a graph homomorphism $\alpha : G \to T$ *into* some graph $T$. When we adopt such a viewpoint, we will call $\alpha$ an *adherence morphism*, and we can think of $T$ as a kind of type graph. For instance, if

$$T \quad = \quad a \;\rightleftarrows\; b$$

then any adherence $\alpha : G \to T$ assigns one of two "colors" $a$ and $b$ to nodes of $G$ such that no two $G$-neighbors have the same color ($\alpha$ is effectively a proof that $G$ is 2-colorable or bipartite); and if

$$T \quad = \quad \circlearrowleft x \circlearrowright$$

then we can regard any $\alpha : G \to T$ as assigning every edge of $G$ one of two "colors", depending on which loop is being assigned. We now consider a ToyPB rule to specify a manipulation of such type graphs.

**Definition 6** (ToyPB Rule). A *ToyPB rule* is a morphism $\rho : L' \leftarrow R'$. $L'$ and $R'$ are called *type graphs*.

Observe that ToyPO and ToyPB rules are formally identical. But it helps to emphasize the difference in perspective by presenting the definitions differently.

*Example* 4. Rule 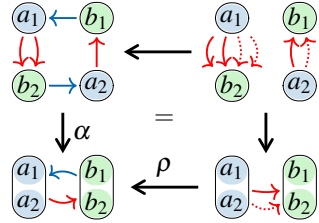 specifies the deletion of edges that originate in $b$-nodes, and the duplication of edges that originate in $a$-nodes. An intended example application is given by the commutative square in Figure 2. In this and the next example, the rule lies at the bottom of the square, and the node identities of the rule have been adjusted to indicate how the morphisms are defined (see Notation 1).

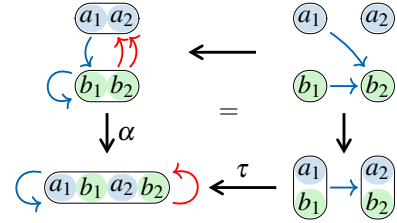

Figure 2: Example 4.          Figure 3: Example 5.

*Example* 5. Let us again think of an adherence $\alpha : G \to T$ on graphs $G$ and $T$ as assigning colors to elements of $G$. Rule (a b) ← (a)→(b) specifies the deletion of all "red" edges, and splits all nodes so that the "blue" edges are rendered as edges of a (one-way) bipartite graph. An intended example application is given by the commutative square in Figure 3.

The desired outcomes of Examples 4 and 5 (which we hope the reader agrees are the most reasonable ones given the adopted intuitive perspective) can be obtained through a pullback construction, a notion that is the dual to the pushout (Definition 3). Like pushouts, pullbacks are unique up to isomorphism.

**Definition 7** (Pullback [20, 2]). The *pullback* of a cospan $G \xrightarrow{\alpha} L' \xleftarrow{\rho} R'$ is a span $\sigma = G \xleftarrow{i_G} H \xrightarrow{i_R} R$ such that

1. $\sigma$ *is a candidate solution*: $\alpha \circ i_G = \rho \circ i_R$; and
2. $\sigma$ *is the universal solution*: for any span $G \xleftarrow{i_G'} H' \xrightarrow{i_R'} R'$ that satisfies $\alpha \circ i_G' = \rho \circ i_R'$, there exists a *unique* morphism $x : H' \to H$ such that $i_G' = i_G \circ x$ and $i_R' = i_R \circ x$.

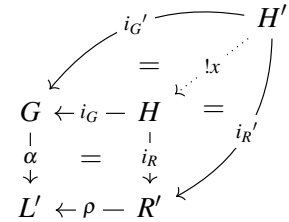

Figure 4: Pullback.

Both requirements are conveniently visualized in the commuting diagram in Figure 4.

*Remark* 2 (Fibered Product). The metaphor of a pushout as a gluing construction unfortunately does not dualize very well. However, in the category of sets, the pullback can be understood as a *fibered product*, which is a generalization of the familiar Cartesian product: the pullback object $H$ (Figure 4) contains all pairs $(x, y)$ ($x \in G$ and $y \in R'$) for which $\alpha(x) = \rho(y)$. For the category of graphs, the vertex set $V$ and edge set $E$ of $H$ can in fact be constructed by constructing the fibered products for $V$ and $E$ independently.

*Exercise* 2. Adapt Exercise 1 for pullbacks: construct some candidates that compete with the solutions given in Examples 4 and 5 (e.g., consider the empty graph), and assess why these candidates do not qualify as pullbacks.

We may now define the notion of a ToyPB rewrite step.

**Definition 8** (ToyPB Rewrite Step)**.** A ToyPB rule $\rho : L' \leftarrow R'$ and adherence morphism $\alpha : G \to L'$ induce a *ToyPB rewrite step* $G \Rightarrow_{\text{ToyPB}}^{\rho,\alpha} H$ if there exists a pullback of the form

$$\begin{array}{ccc} G & \leftarrow i_G - & H \\ \downarrow \alpha & \text{PB} & \downarrow i_R \\ L' & \leftarrow \rho - & R' \end{array}.$$

Note that unlike ToyPO, ToyPB does not have any injectivity restrictions, because we would usually like arbitrarily large graphs $G$ to be typeable by a relatively small type graph $L'$.

## 4 PBPO$^+$

We can now almost introduce PBPO$^+$. The following preliminary exercise shows that we can use pullbacks to construct preimages. It is useful for understanding PBPO$^+$ rules and matches.

*Exercise* 3 (Computing Preimages)**.** For cospans $L \xrightarrow{t_L} L' \xleftarrow{l'} K'$ with one injective leg $t_L$, define $l'^{-1}(t_L)$ as the subgraph of $K'$ mapped onto the subgraph $t_L(L)$ of $L'$. Using intuitive reasoning, convince yourself that for a pullback

$$\begin{array}{ccc} L & \leftarrow l - & K \\ \hookrightarrow t_L & \text{PB} & \downarrow t_K \\ L' & \leftarrow l' - & K' \end{array},$$

we have $K \cong l'^{-1}(t_L)$.

**Definition 9** (PBPO$^+$ Rewrite Rule [7, 19])**.** A *PBPO$^+$ rewrite rule* $\rho$ is a diagram of the form

$$\rho \;=\; \begin{array}{ccccc} L & \leftarrow l - & K & - r \to & R \\ \downarrow t_L & \text{PB} & \downarrow t_K & & \\ L' & \leftarrow l' - & K' & & \end{array}$$

where $L$ is the *lhs pattern* of the rule, $L'$ its *(context) type*, and $t_L$ its *(context) typing*. Likewise for the *interface K*. $R$ is the *rhs pattern* or *replacement for L*.

For injective $t_L$ (which we assume in this paper), one can think of $L'$ as the type graph of ToyPB, with $t_L : L \to L'$ an embedding that distinguishes a pattern $L$ in $L'$. The morphism $l' : L' \leftarrow K'$ is used to specify deletion and duplication on the type graph, similar to a ToyPB rule. A morphism $r : K \to R$ is then used to specify identification and addition specifically on $l'^{-1}(t_L)$, similar to a ToyPO rule.

For PBPO$^+$ matches, we would like to view $L'$ as providing a context typing for the designated pattern $t_L(L) \subseteq L'$. For rewrite steps, this means that *precisely one* occurrence of $L$ in $G$ should be mapped onto $t_L(L)$ by an adherence morphism $\alpha : G \to L'$. Stated in terms of preimages, we require that $\alpha^{-1}(t_L) \cong L$. We call such a match a *strong match*.

**Definition 10** (Strong Match [19])**.** An *adherence morphism* $\alpha : G \to L'$ establishes a *strong match* for a context typing $t_L : L \to L'$ if the square

$$\begin{array}{ccc} L & - m \to & G \\ \| & \text{PB} & \downarrow \alpha \\ L & - t_L \to & L' \end{array}$$

is a pullback square. The induced morphism $m : L \to G$ is called the *match morphism*.

Because we restrict to injective $t_L$, the strong match implies that match morphisms $m$ are injective.[1]

*Example* 6. The left of

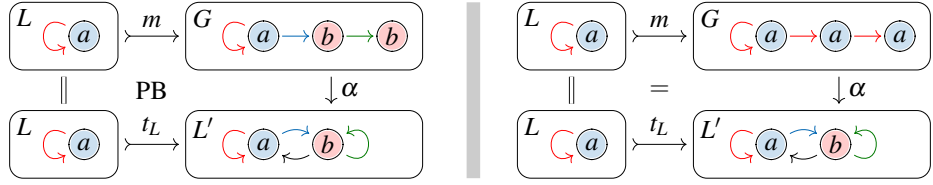

visualizes a strong match: $L'$ has exactly one occurrence of $L$ (namely $m(L)$) mapped onto $t_L(L)$. The square is a pullback square. The right is merely a commutative square, not a pullback square: because all of $G$ is collapsed onto $t_L(L)$, the pullback object $\alpha^{-1}(t_L)$ is in fact $G$ itself.

The definition of a PBPO$^+$ rewrite step consists of a strong match square, a ToyPB-like pullback square (using "rule" $l' : L' \leftarrow K'$), and a ToyPO-like pushout square (using "rule" $r : K \to R$). For this, the pullback and pushout squares need to be appropriately connected. As stated in the definition below, this can always be done, because the existence of a unique morphism $u : K \to G_K$ satisfying $t_K = u' \circ u$ is guaranteed. It can in fact be obtained by pulling $t_K$ along $u'$ [19, Lemma 15].

**Definition 11** (PBPO$^+$ Rewrite Step [19]). A PBPO$^+$ rewrite rule $\rho$ (Definition 9), match morphism $m : L \to G_L$ and adherence morphism $\alpha : G_L \to L'$ induce a rewrite step $G_L \Rightarrow^{\rho,(m,\alpha)}_{\text{PBPO}^+} G_R$ if the properties indicated by the commuting diagram

$$\begin{array}{ccccccc}
 & & & & K & \xrightarrow{\ r\ } & R \\
 & & & \nearrow {\scriptstyle !u} & \downarrow & \text{PO} & \downarrow w \\
L & \xrightarrow{m} & G_L & \xleftarrow{g_L} & G_K & \xrightarrow{g_R} & G_R \\
\| & \text{PB} & \downarrow \alpha & \text{PB} & {\scriptstyle u'}\downarrow \swarrow {\scriptstyle t_K} & & \\
L & \xrightarrow{t_L} & L' & \xleftarrow{l'} & K' & & \\
\end{array}$$

hold, where $u : K \to G_K$ is the unique morphism satisfying $t_K = u' \circ u$ [19, Lemma 15].[2]

**Proposition 4.1** ([19, Lemma 15]). *If $t_L$ or $m$ is injective, $u$ is injective.* □

*Example* 7 (Rewrite Step). The rewrite step $G_L \Rightarrow^{\rho,(m,\alpha)}_{\text{PBPO}^+} G_R$ depicted in Figure 5 illustrates some important features of PBPO$^+$. The rule $\rho$ consists of the objects $L, K, R, L'$ and $K'$ together with the obvious morphisms. The pattern $L$ requires any host graph $G_L$ to contain three nodes, and two of these nodes have an edge targeting the third node. The graph $L'$ describes the permitted shapes of the host graph $G_L$ around the pattern $m(L)$. Due to the strong match condition, any $\alpha : G_L \to L'$ has to map all nodes and edges in the context $G_L - m(L)$ onto $L' - t_L(L)$. So, in particular, $c_1 c_2$ in $L'$ captures all the context nodes of $G_L$. Moreover, each edge in $L' - t_L(L)$ is a placeholder for an arbitrary number (zero or more) of edges in the host graph. This example illustrates the following features:

(i) *Application conditions:*

The graph $L'$ allows for an arbitrary number of edges from $x_1 x_2$ to the context, from the context to $z$, from $y_1 y_2$ to $x_1 x_2$, and from $y_1 y_2$ to $z$. Due to the loop on $c_1 c_2$, any edges among context nodes are allowed as well. Apart from these permitted edges and the required edges in the pattern, any other edges are forbidden in $G_L$. For instance, there cannot be edges from $z$ to the context, and no additional edges from $x_1x_2$ to $y_1y_2$.

---

[1]This follows immediately on a categorical level, using what is called "pullback stability of monomorphisms".

[2]Morphism $u$ can be obtained by pulling back $m$ along $g_L$. However, as shown in the cited lemma, the solution for $u$ in $t_K = u' \circ u$ exists and is uniquely determined for PBPO$^+$ (it is not for PBPO [19, Remark 21]). We thus need not specify how $u$ is constructed, allowing us to simplify the diagram.

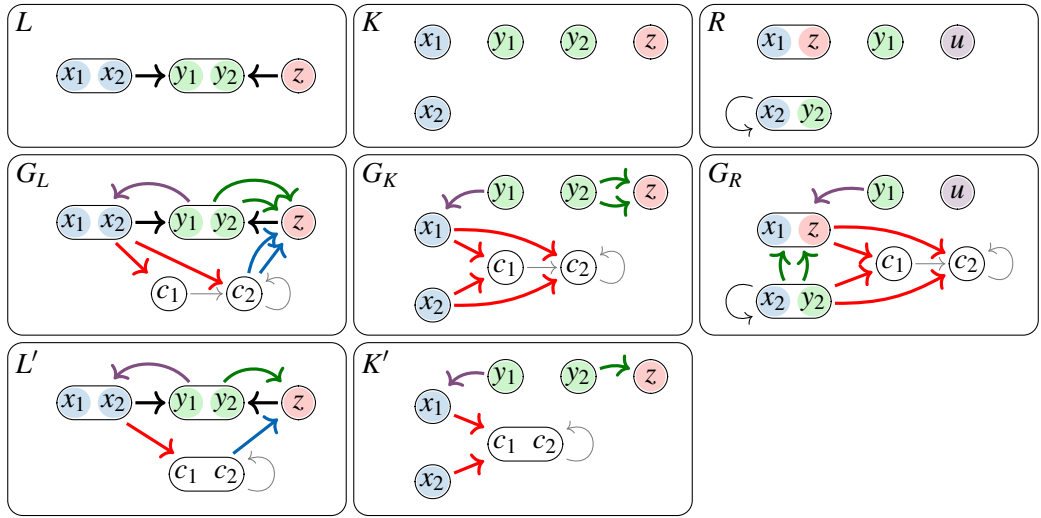

Figure 5: Example illustrating duplication, deletion and redirection and constraining application.

(ii) *Duplicating and deleting elements:*

The morphism $l' : K' \to L'$ enables the duplication and deletion of nodes and edges. For instance, from $L'$ to $K'$, the node $x_1x_2$ is duplicated along with its edges to the context (indicated in red). The edges from $c_2$ to $z$ and the pattern edges are deleted, because they do not lie in the image of $l'$. Both transformations occur along $g_L : G_K \to G_L$ as well.

(iii) *Identifying and adding elements:*

The morphism $r : K \to R$ enables the identification of elements of $K$, and the addition of new elements. Here, $K$ is the result of restricting the duplication and deletion effects of $l'$ to $t_L(L) \cong L$. In the example, $r$ identifies (or merges) $x_1$ with $z$ and $x_2$ with $y_2$, adds a fresh node $u$, and a fresh edge from $x_2y_2$ to itself. In the middle row, observe that the sources and targets of edges are updated accordingly from $G_K$ to $G_R$. For instance, the edges $y_1 \to x_1$ and $y_2 \to z$ in $G_K$ are redirected to target the merged node $x_1z$ in $G_R$.

(iv) *Edge redirection:*

The combination of duplication along $l' : K' \to L'$ and merging along $r : K \to R$ enables the arbitrary redirection of all those endpoints (sources or targets) of edges that lie in the pattern $m(L)$. Importantly, the edge itself does not need to be part of the pattern, only the endpoint to be redirected. For instance, the edges from $y_1y_2$ to $z$ (indicated in green) are redirected to go from $x_2$ to $x_1$. This is achieved by first duplicating one endpoint of the edge, namely the source $y_1y_2$, and then merging the fresh source $y_2$ with $x_2$, and the target $z$ with $x_1$.

For another example, see [19, Example 17].

## 5 Modeling Binary Decision Diagrams

So far we have exclusively considered unlabeled directed multigraphs. But in many application domains, graph edges and/or nodes are labeled. More formally, a *node and edge labeled directed multigraph* or simply *labeled graph* is a graph $G = (V, E, s, t)$ equipped with two labeling functions $\ell^V : V \to \mathscr{L}$ and $\ell^E : E \to \mathscr{L}$ for some fixed set of labels $\mathscr{L}$.

When considering these labeled graphs as the objects of a category, one must specify what are the morphisms between them. The default definition of a labeled graph morphism $\phi : G \to H$ is a homomorphism on the unlabeled graphs underlying $G$ and $H$ that moreover respects the labelings, i.e., $\ell^V_H \circ \phi = \ell^V_G$ and $\ell^E_H \circ \phi = \ell^E_G$. For PBPO$^+$ we introduced a relaxation of this criterion (roughly, $\leq$ instead of $=$ in the equations) that provides a lot of convenience in defining context type graphs (such as $L'$), as well as flexible support for relabeling. The corresponding category is called $\mathbf{Graph}^{(\mathscr{L}, \leq)}$ (where $(\mathscr{L}, \leq)$ is a complete lattice, to be explained below).

In this section, we provide an example of how PBPO$^+$ coupled with $\mathbf{Graph}^{(\mathscr{L}, \leq)}$ can be used to specify the reduction rules for binary decision diagrams (BDDs) [15, Chapter 6]. BDDs are familiar data structures, used to represent Boolean functions efficiently. For instance, the Boolean function $f(p, q) = p \wedge q$ admits BDD representations

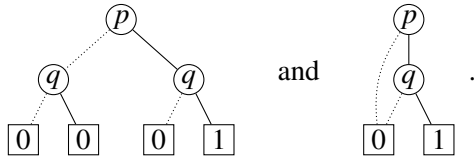

To evaluate $f(p, q)$, given some assignment of truth values to $p$ and $q$, start at the root node of a BDD representation and walk down the tree as follows:

- if you are at an internal node, it is labeled with some variable $x$: follow the dotted line if $x$ is false, and the solid line if $x$ is true; and
- if you are at a leaf node, read off the evaluation of $f(p, q)$, where 1 is true and 0 is false.

Of the two given BDDs for $f(p, q)$, the left tree is essentially the truth table representation of $f(p, q)$, and its size is exponential in the number of variables. The tree on the right is a more efficient (and in fact optimal) representation. It results by applying reduction rules to the tree on the left, which merge repeated substructures and eliminate vacuous decisions. In what follows, we will model BDDs as objects of $\mathbf{Graph}^{(\mathscr{L}, \leq)}$, and implement BDD reduction rules with PBPO$^+$ rules.

First, we require a specific type of order on the set of labels.

**Definition 12** (Complete Lattice). A *complete lattice* is a partially ordered set (poset) $(\mathscr{L}, \leq)$ such that all subsets $S$ of $\mathscr{L}$ have a supremum (join) $\bigvee S$ and an infimum (meet) $\bigwedge S$.

A complete lattice can be viewed as a category. The objects of this category are the elements of $\mathscr{L}$, and a (unique) arrow $x \to y$ exists iff $x \leq y$. Importantly, pushouts and pullbacks exist, and correspond to computing joins and meets, respectively. Every complete lattice moreover has a top element $\top$ and a bottom element $\bot$. Examples of complete lattices include: $(\{S' \mid S' \subseteq S\}, \subseteq)$ for a set $S$; total orders that have a minimum and a maximum; as well as certain type lattices, such as the following definition.

**Definition 13** (BDD Lattice). Let $\mathscr{X} = \{x_1, \ldots, x_n\}$ be a finite set of variables, and $\mathbf{Bool} = \{0, 1\}$ a set of truth values. Let $\mathscr{L}_{\text{BDD}} = \mathscr{X} \uplus \mathbf{Bool} \uplus \{\mathscr{X}, \mathbf{Bool}\} \uplus \{\top, \bot\}$ be a set of labels. We define the *binary decision diagram complete lattice* $(\mathscr{L}_{\text{BDD}}, \leq)$ as given by the diagram

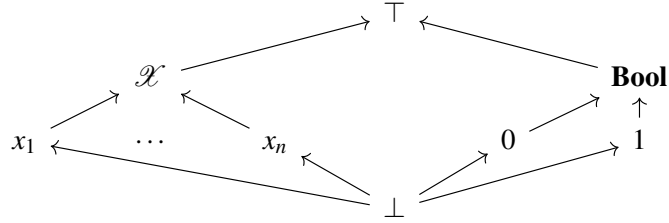

*Example* 8 (Rewriting in $(\mathscr{L}_{\text{BDD}}, \leq)$). Assume $x_1, x_2 \in \mathscr{X}$. The two rules and (examples of) steps

$$L : \bot \xleftarrow{l} K : \bot \xrightarrow{r} R : x_1$$
$$\downarrow m \quad \text{PB} \quad \downarrow u \quad \text{PO} \quad \downarrow w$$
$$G_L : x_2 \xleftarrow{g_L} G_K : \bot \xrightarrow{g_1} G_R : x_1$$
$$\downarrow \alpha \quad \text{PB} \quad \downarrow u'$$
$$L' : \mathscr{X} \xleftarrow{l'} K' : \bot$$

and

$$L : \bot \xleftarrow{l} K : \bot \xrightarrow{r} R : \bot$$
$$\downarrow m \quad \text{PB} \quad \downarrow u \quad \text{PO} \quad \downarrow w$$
$$G_L : 0 \xleftarrow{g_L} G_K : 0 \xrightarrow{g_R} G_R : 0$$
$$\downarrow \alpha \quad \text{PB} \quad \downarrow u'$$
$$L' : \top \xleftarrow{l'} K' : \top$$

show how PBPO$^+$ rewriting behaves in the complete lattice category $(\mathscr{L}_{\text{BDD}}, \leq)$. The rule on the left allows matching any element $\bot \leq x \leq \mathscr{X}$, and replaces $x$ with $\bigvee\{\bigwedge\{x, \bot\}, x_1\} = \bigvee\{\bot, x_1\} = x_1$. Assuming $\bot$ and $\mathscr{X}$ do not occur in the domain of rewritable objects, the rule thus specifies the replacement of an arbitrary variable $x \in \mathscr{X}$ with the fixed variable $x_1$ (in the example, $x = x_2$). The rule on the right allows matching any element $y$, and replaces it with $\bigvee\{\bigwedge\{y, \top\}, \bot\} = \bigvee\{y, \bot\} = y$. In other words, the rule does not restrict matches, and does not modify matches. This behavior is useful in the labeled graph setting, where we usually want that parts of the graph (in particular the context) do not influence, and stay stable under, rewriting.

We use the following category for rewriting BDDs, where the label set is instantiated with $(\mathscr{L}_{\text{BDD}}, \leq)$.

**Definition 14 (Graph$^{(\mathscr{L}, \leq)}$ [18]).** For a complete lattice $(\mathscr{L}, \leq)$, the category **Graph**$^{(\mathscr{L}, \leq)}$ has graphs labeled from $\mathscr{L}$ as objects, and arrows are graph homomorphisms $\phi : G \to G'$ that satisfy $\ell_G(x) \leq \ell_{G'}(\phi(x))$ for all $x \in V_G \cup E_G$.

The pullbacks and pushouts in **Graph**$^{(\mathscr{L}, \leq)}$ exist. Given a cospan $B \to D \leftarrow C$ of labeled graphs, the pullback $B \xleftarrow{f} A \xrightarrow{g} C$ is obtained by constructing the pullback on the underlying unlabeled graphs and labeling each element $x \in V_A \cup E_A$ with the meet $\bigwedge\{\ell(f(x)), \ell(g(x))\}$. Dually for pushouts.

**Definition 15 (Binary Decision Diagram [5]).** A *binary decision diagram* (*BDD*) $D$ is a labeled directed graph $(V, E, s, t, \ell^V : V \to \mathscr{L}_{\text{BDD}}, \ell^E : E \to \mathscr{L}_{\text{BDD}})$ where

1. the graph has a single root;
2. all leaves and edges are labeled from **Bool**;
3. all internal nodes are labeled from $\mathscr{X}$;
4. every internal node has two outgoing edges, one labeled with 0 and one labeled with 1; and
5. variables are not repeated along paths.

A BDD is *reduced* [5, Definition 5] if

1. it does not contain two distinct isomorphic subgraphs; and
2. no node has its two outgoing edges target the same child.

**Definition 16 (BDD Reduction Rules with PBPO$^+$).** For the visual representation of rules, we let edges with arrowheads at both ends represent two edges (one in each direction). The PBPO$^+$ rules to implement BDD reduction ($|\mathscr{X}| + 3$ in total) can then be given as follows:

- Identically labeled leaves are pairwise identified. One rule LEAF$_b$ for every $b \in \mathbf{Bool} = \{0, 1\}$, as captured by the schema:

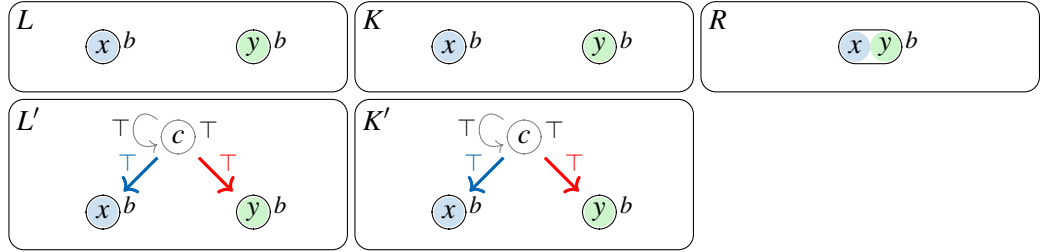

- Identically labeled internal nodes whose decisions lead to the same subtrees are pairwise identified. One rule MERGE-ISO$_{x_i}$ for every $x_i \in \mathscr{X}$, as captured by the schema:

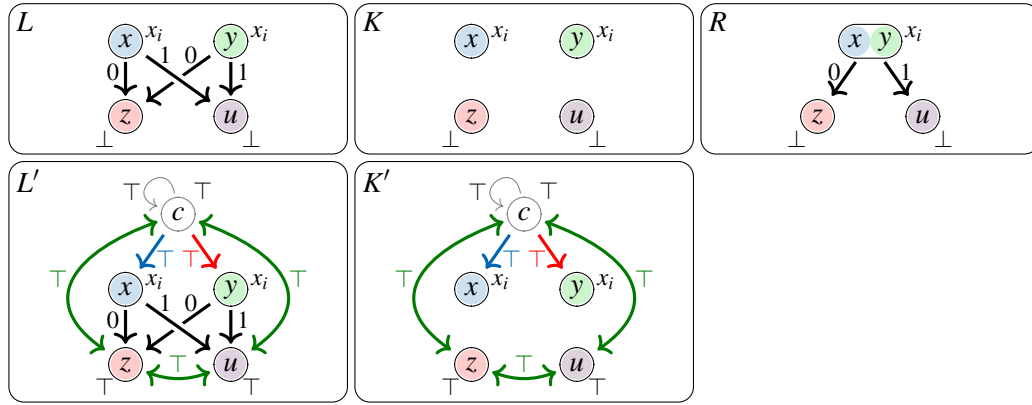

- Vacuous decisions are removed. That is, if both outgoing edges of an internal node $x$ target the same node $y$, then $x$ is removed, and all of $x$'s incoming edges are redirected to $y$. This is captured by the single rule ELIM-VACUOUS (an example application is given in the middle row):

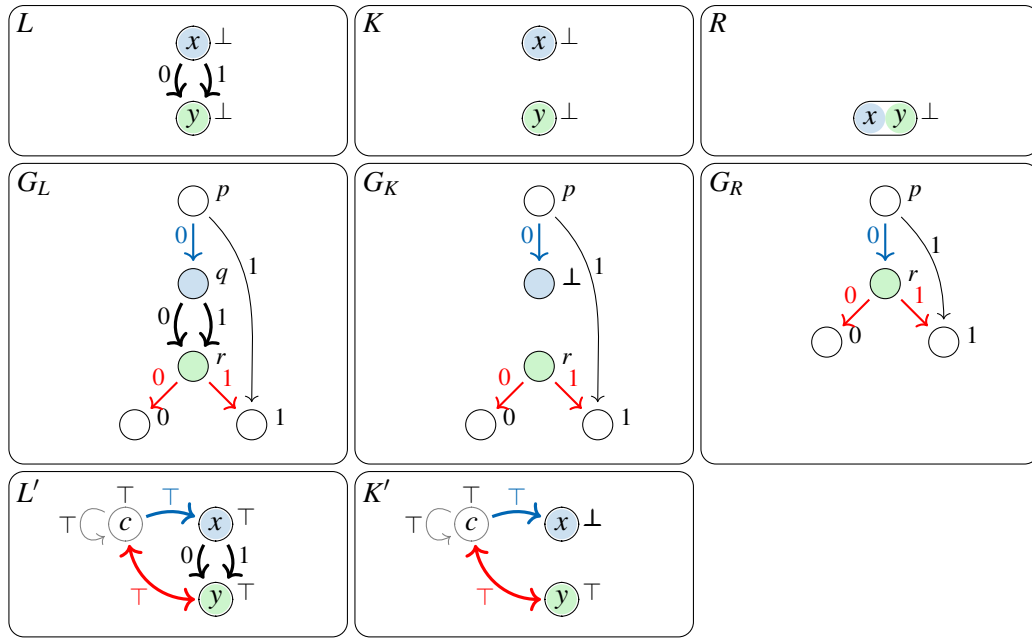

*Remark* 3. The rule schemas are used because equality of labels cannot be expressed in the base formalism. Of course, support for such constraints could be added to reduce the number of rules.

*Example* 9 (BDD Reduction). The sequence

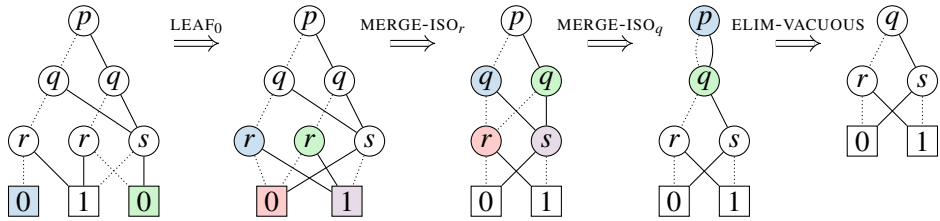

is an example of a BDD reduction. The colors match the pattern colors of the corresponding rule. Observe that the subgraphs rooted at the two $q$ nodes were isomorphic from the start, and that the rules implement a bottom-up strategy for merging isomorphic subgraphs.

**Lemma 5.1** (Correctness of BDD Reduction). *The BDD reduction rules of Definition 15 terminate. Moreover, if the reduction rules are maximally applied to a BDD, then the result is a reduced BDD.*

*Proof.* The rules terminate because each rule reduces the number of vertices by one.

To show that the result is a reduced BDD, we first argue that the BDD properties are invariant. We then argue that a rewrite rule is applicable as long as one of the reduced BDD criteria is violated. Using the termination result, it then follows that the end result is a reduced BDD.

For invariance of the BDD properties, we consider each property in turn:

1. *The graph has a single root*: None of the rules create new nodes. Schema MERGE-ISO$_{x_i}$ reduces the in-degree of $z$ and $u$, but they do not become roots. Rule ELIM-VACUOUS effectively deletes $x$, and $y$ inherits the in-degree of $x$. All other in-degrees are preserved.
2. *All leaves and edges are labeled from* **Bool**: First, none of the rules create nodes, and the edges added by schema MERGE-ISO$_{x_i}$ are correctly labeled. Second, for existing elements, node and edge labels in the context are fully preserved by each rule, due to the use of identical labels for these elements in $L'$ and $K'$. For pattern elements, only ELIM-VACUOUS changes a label (the label of $x$ is effectively erased), but this node is then merged with $y$, whose label is preserved.
3. *All internal nodes are labeled from* $\mathscr{X}$: The reasoning for the previous item applies.
4. *Every internal node has two outgoing edges, one labeled with* 0 *and one labeled with* 1: For any nodes in the context of each rule, the outgoing edges are preserved. The pattern nodes $x$ and $y$ of MERGE-ISO$_{x_i}$ are merged and assigned the correct pair of outgoing edges. The pattern node $x$ of ELIM-VACUOUS is effectively deleted. All other pattern nodes preserve their outgoing edges.
5. *Variables are not repeated along paths*: Schemas LEAF$_b$ and MERGE-ISO$_{x_i}$ merge isomorphic nodes and hence do not change the sequences of nodes along paths. Rule ELIM-VACUOUS effectively deletes $x$, strictly shrinking some paths.

It remains to show that if a BDD is not a reduced BDD, then a rewrite rule applies. Observe that if the second criterion for being a reduced BDD is violated, rule ELIM-VACUOUS applies. For the case where the first property is violated, we argue that for any two distinct isomorphic subgraphs rooted in $u$ and $v$, respectively, at least one rule applies. We proceed by induction on the depth $d$ of these subgraphs:

- If $d = 0$, then $u$ and $v$ are distinct leaves labeled with $b \in$ **Bool**, so that rule LEAF$_b$ can be applied.
- For the induction step, assume $d = n+1$, and let $0(x)$ and $1(x)$ denote the 0-child and 1-child of internal nodes $x$, respectively. Observe that the trees rooted in $0(u)$ and $0(v)$ are isomorphic and have depth $n$; and likewise for $1(u)$ and $1(v)$. So if $0(u) \neq 0(v)$ or $1(u) \neq 1(v)$, by the induction hypothesis a rule can be applied. Moreover, if $0(u) = 1(u)$ or $0(v) = 1(v)$, ELIM-VACUOUS can be applied. This leaves the case where $0(u) = 0(v)$, $1(u) = 1(v)$, $0(u) \neq 1(u)$ and $0(v) \neq 1(v)$. Here, rule MERGE-ISO$_{x_i}$ can be applied, for $x_i$ the label of $v$ and $u$.   □

*Remark* 4. We are currently developing a termination method for PBPO$^+$ rewrite systems. Using this method, proving termination of this particular BDD reduction system is trivial. It would be interesting to also develop methods for verifying invariance properties.

*Remark* 5. Strictly speaking, it is possible to define the BDD reduction rules in the ordinary category of node- and edge-labeled graphs, where the homomorphisms $\phi : G \to G'$ are label-preserving ($\ell_G(x) = \ell_{G'}(\phi(x))$). But this results in an increase of rules exponential in the number of variables $\mathscr{X}$. For instance, to model rule ELIM-VACUOUS alone, one needs a rule for every possible pair of labels on vertices $x$ and $y$. Moreover, for each rule, the context type graphs also explode. For instance, for the $L'$ of ELIM-VACUOUS, the $\top$-labeled context node $c$ must be replaced by $|\mathscr{X}|$ nodes $C$, each node $c \in C$ with a unique label $x_i \in \mathscr{X}$; and instead of the four $\top$-labeled edges (observe that the red edge between $c$ and $y$ is bidirectional), one must add edges

$$\bigcup \{ \{c_1 \xrightarrow{b} c_2, \ c_1 \xrightarrow{b} x, \ c_1 \xrightarrow{b} y, \ y \xrightarrow{b} c_1\} \mid c_1, c_2 \in C, \ b \in \mathbf{Bool} \}$$

to graph $L'$.

## 6  Comparison with Double Pullback Rewriting

As has been kindly pointed out by anonymous reviewers, PBPO$^+$ bears some similarity to the double pullback rewriting approach of Bauderon [3, 4], an approach approximately dual to double pushout rewriting. We compare the two in this section.

    Bauderon's approach in [3] is slightly different from the approach by Bauderon and Jacquet in [4]. We focus mainly on the former, and comment on the latter. The following definition captures the essence of any double pullback approach.

**Definition 17** (Double Pullback Rewriting [3]). A *double pullback rule* $\rho$ is of the form $L \xrightarrow{l} A \xleftarrow{r} R$. An *occurrence* is a morphism $m : G \to L$. There exists a *double pullback rewrite step* from object $G$ to object $H$, induced by rule $\rho$ at occurrence $m : G \to L$, if there exists a diagram of the form

$$\begin{array}{ccccc}
G & \xrightarrow{q} & D & \longleftarrow & H \\
{\scriptstyle m}\downarrow & \text{PB} & {\scriptstyle u}\downarrow & \text{PB} & \downarrow \\
L & \xrightarrow{l} & A & \xleftarrow{r} & R
\end{array}$$

where $G \xrightarrow{q} D \xrightarrow{u} A$ is called a *pullback complement* for $G \xrightarrow{m} L \xrightarrow{l} A$.

    A *double pullback rewriting* (*DPU*) *approach for a class of categories* $\mathscr{C}$ adds constraints on rules, occurrences and steps for a designated class of categories $\mathscr{C}$.

    Before we discuss Bauderon's approach in [3] in more detail, we provide an intuition for the general idea. Recall that a pullback of a cospan $A \to 1 \leftarrow B$ with 1 the terminal object (in **Set** this is a singleton set) is just the product $A \times B$ (the cartesian product in **Set**). And recall that 1 is a unit for the product. This means that for any $A$, the pullback object of $A \to 1 \leftarrow 1$ is $A$.

    We illustrate how this observation can be used by means of an example in **Set**. Suppose that we wish to define a rule that replaces a 3-element subset by a 2-element subset (up to isomorphism, then, one element is deleted from the superset). The bottom cospan of diagram

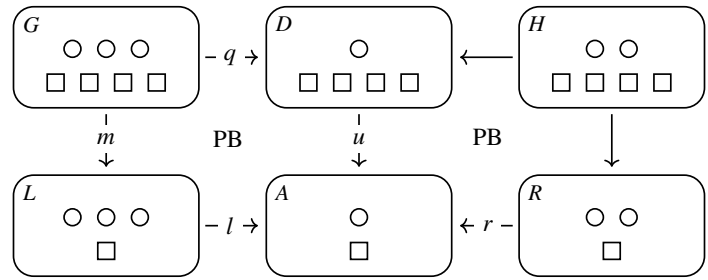

depicts a rule $\rho = L \xrightarrow{l} A \xleftarrow{r} R$ that implements our specification. The middle set $A$ contains two elements, which are terminal subobjects in **Set** (informally: sinks), one depicted as a circle and the other as a square. We will let the circle of $A$ function as the sink for pattern elements (the elements of $G$ that are to be replaced), and the square of $A$ as the sink for context elements (the remaining elements of $G$ that must be preserved). The circles and boxes in $L$ indicate how the four elements of $L$ are mapped: three elements are mapped into the sink for pattern elements, and there is a copy of the context sink. Likewise for $R$.

The top cospan depicts an application of rule $\rho$ to set $G$ for occurrence $m$, rewriting it to set $H$. Along the vertical arrows, the circles are mapped injectively.

For the left pullback, the square in $L$ acts as a neutral element for the square elements, with the effect that the squares of $G$ are preserved isomorphically into $D$ along $q$. Similarly, for $G$ to qualify to as a pullback object, the subset of circles in $D$ must in this example form a unit, i.e., a singleton.

For the right pullback, the context of $D$ is reflected isomorphically into $H$ because $R$ is neutral for the context elements. Symmetrically, the right-hand side pattern of $R$ is reflected isomorphically into $H$ because $D$ is neutral for the pattern elements.

The depicted rule and step satisfy the following criteria: any context is allowed by the rule, the context of $G$ is isomorphically preserved into $H$, the pattern of $G$ is isomorphic to the pattern of $L$, and the pattern of $L$ is substituted for the pattern of $R$ in $G$. Some formal constraints on the rule, occurrence and step would be needed to ensure that this holds for all rules and steps. But we need not concern ourselves with the details here.

Now, for a graph-like category the idea is similar: we at least have two sinks in $A$ (two terminal graphs: i.e., two nodes with a loop), one as a sink of the pattern graph, and one as the sink for the context graph. But the main difference is that in principle there may be additional edges between the pattern and the context in $G$, and on the pattern itself. The formal constraints of the formalism must be adapted to accommodate that to a desired extent.

Bauderon's first version [3] is defined specifically for **SimpleGraph**, the category of simple directed graphs. That is, the objects are tuples $(V, E)$ with $E \subseteq V \times V$, and the morphisms are the usual structure-preserving ones.[3] Observe that in simple graphs, parallel edges are not possible.

Bauderon defines an *alphabet graph A* as the graph obtained by taking the complete and reflexive graph on node set $\{-1\} \cup \mathbb{N}$, and then removing the edge between 0 and $-1$ [3, Definition 6]. Intuitively, $-1$ and its loop represent what we call the pattern sink, 0 and its loop represent a sink for all context nodes and edges not directly neighbouring the pattern (hence the deletion of the edge between $-1$ and 0), and all remaining nodes of $A$ represent nodes not part of the pattern, but connected to it. Bauderon calls these *interface items*.

A *VR-rule* is a morphism $x : X \to A$ such that $|x^{-1}(0)| = 1$ and for all $i \in \mathbb{N}_{>0}$, $|x(i)^{-1}| \leq 1$ [3, Definition 8]. So $X$ has exactly one context sink, and $x$ is injective on interface items.

---

[3]In the running text, Bauderon writes that he considers undirected graphs. Graphs in figures also appear to be undirected. But mathematically, the graphs are clearly directed. In any case, the essence of our observations hold for either interpretation.

The constraint on occurrences is incorrectly defined in [3, Definition 15], but following Bauderon's explanatory text it can be stated as follows. Let $p : 1 \rightarrowtail A$ be the injective map (where the domain denotes the terminal graph) that selects $-1$ and its loop (the pattern sink) in $A$. For occurrences $m : G \to L$ and VR-rules $l : L \to A$, Bauderon requires that

$$\begin{array}{ccccc}
P & = & P & \longrightarrow & 1 \\
{\scriptstyle p'}\downarrow & & {\scriptstyle mp'}\downarrow & & \downarrow{\scriptstyle p} \\
 & \text{PB} & & \text{PB} & \\
G & \xrightarrow{m} & L & \xrightarrow{l} & A
\end{array}$$

holds for some $P$ and $p' : P \rightarrowtail G$. In other words, $m$ maps precisely one subgraph of $G$ isomorphic to $P$ onto the designated pattern $P$ of $L$, and the remaining elements of $G$ onto the other nodes of $L$.

Bauderon proves that for VR-rules $l : L \to A$ and constrained occurrences $m : G \to L$, the existence of a pullback complement is guaranteed [3, Proposition 16].

A DPU rewrite rule is finally defined as a cospan $L \xrightarrow{l} A \xleftarrow{r} R$ [3, Definition 17].

We can now make three observations in relation to PBPO$^+$:

1. Pullback complements are not always unique using the definition of DPU in [3], sometimes causing nondeterminism even for basic rules that would be properly implementable with PBPO$^+$.

   For an example, consider the diagram

   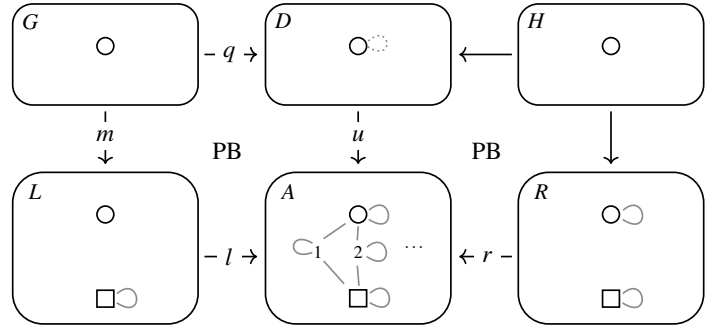

   which depicts a rule and step, where pattern sink $-1$ is represented by a circle and context sink $0$ is represented by a square in $A$. Every depicted edge between nodes $x$ and $y$ represents edges $(x,y)$ and $(y,x)$ (so it represents one edge if $x = y$, and two edges otherwise).

   The rule (bottom span) specifies the addition of a loop to a graph component that is just a single node (the circle). The application (top span) applies the rule to the singleton node graph. Two pullback complements exist: the node in $D$ can either have a loop or not. If it does not have a loop, the depicted $H$ results, which would constitute an identity step. Only if the node in $D$ has a loop, the circle in $H$ would have the intended loop.

   The general observation (valid also for other graph-like categories) is that, if $l$ is not at least surjective on the pattern sink, then the pattern sink may not be reflected into $D$, possibly causing a loss of elements of the right hand side pattern [4, Proposition 4]. A constraint on the pullback complement would therefore also be needed.

2. In PBPO$^+$ (as well as DPO), the middle interface graph $K$ of a rule can represent a complex subobject of $L$, that is reinserted into the right hand side $R$. For DPU, as presented here, such explicit mediation is not possible, because the pattern on the left is collapsed into a terminal object. This is a limitation for some categories.

   For instance, consider category $\mathbf{Graph}^{(\mathscr{L}, \leq)}$, where the graphs $L$ and $L'$ of a PBPO$^+$ rule work together to specify lower and upper bounds on labels of patterns, respectively. For any pattern instance that satisfies these bounds, one can specify the preservation of some subobject of the pattern instance into the right side. With DPU, we do not see how a subobject of a match could be preserved into the right, due to the lack of explicit mediation.

3. DPU, as defined by Bauderon in [3], is defined for the very specific notion of directed simple graphs. And to obtain a deterministic rewriting behaviour (arguably the main criterion for well-behavedness), a characterization of the uniqueness of pullback complements is essential. Thus, a generalization (or even translation into another category) must provide a good definition for the alphabet graph $A$, and a good requirement on the pullback complement (see item 1 above). This is non-trivial. By contrast, PBPO$^+$ is defined for any category, and a general sufficient condition for determinism exists for the general setting of quasitoposes [19, Theorem 37].

Let us close with some comments on the version of DPU defined by Bauderon in [4]. This is defined for categories of *G-structured graphs*, which are slice categories of the form **SimpleGraph**$/G$ for a $G \in \text{Obj}(\textbf{SimpleGraph})$. So objects in particular are morphisms $f : D \to G$ into $G$. If $G = 1$, then **SimpleGraph**$/1 \cong$ **SimpleGraph**. If 2 is the graph consisting of two nodes $v$ and $e$ and one edge between them, then **SimpleGraph**$/2$ is isomorphic to the category of hypergraphs [4], where morphisms do not preserve arities of hyperedges. This approach is more general than **SimpleGraph**, but it still fails to cover important categories such as the category of directed multigraphs. The alphabet graph also becomes considerably more complicated. Non-uniqueness of pullback complements moreover remains an issue. Finally, the requirement on occurrences $m : G \to L$ is weakened: any $m$ for which there exists a pullback complement is now considered an occurrence. Among others this means that in any category with strict initial objects (i.e., for all $f : X \to 0$, $X \cong 0$), any rule rewrites the initial object 0 to 0. We do not consider this an improvement over the stronger requirement of [3].

## 7  Related Tutorials

One of the first tutorials by Ehrig et al. [11] present intuitive approaches to DPO and SPO, also covering some metatheoretic properties. Baresi et al. [1] take a different approach and provide a broad and applied introduction to the graph transformation research field, introducing DPO set theoretically. The tutorial by Heckel [14] discusses a notion of typed graph transformation informally. The problem of dangling edges is highlighted, and some solutions that have been proposed to deal with them are discussed. The very recent tutorial by König et al. [16] explains the "essence" of DPO, and gives a gentle build-up towards pushouts. In their case, this involves giving a set-theoretic definition of pushouts. Only SPO is mentioned when the subject of deletion in unknown contexts is discussed. In addition, the tutorial discusses attributed graph rewriting and tools.

For a more extensive overview of tutorials, see [16, Section 7.1].

**Acknowledgments**

We thank Jasmin Blanchette, Wouter Brozius and Femke van Raamsdonk for discussions and feedback. We also thank the anonymous reviewers for their helpful suggestions. Both authors received funding from the Netherlands Organization for Scientific Research (NWO) under the Innovational Research Incentives Scheme Vidi (project. No. VI.Vidi.192.004).